\documentclass[
reprint,
superscriptaddress,
amsmath,amssymb,
aps,
pra,
doi=true,
floatfix,
]{revtex4-2}

\usepackage[colorlinks=true,linkcolor=blue,urlcolor=blue,citecolor=blue]{hyperref}
\usepackage{graphicx}
\usepackage{dcolumn}
\usepackage{bm,upgreek}
\usepackage{xfrac}
\usepackage{xcolor}
\usepackage{mathtools}
\usepackage{amsmath}
\usepackage{CJK}
\usepackage{gensymb}
\usepackage{pifont}
\usepackage{upgreek}
\usepackage[mathscr]{euscript}
\usepackage{makecell}
\usepackage{relsize}
\usepackage[normalem]{ulem}

\begin{document}
\begin{CJK*}{UTF8}{}
\title{A low-crosstalk optical addressing system for atomic qubits based on multiple objectives and acousto-optic deflectors} 

\author{Yi-Long Chen \CJKfamily{gbsn}(陈一龙)}
\affiliation{ 
  CAS Key Laboratory of Quantum Information, University of Science and Technology of China, Hefei 230026, China
}%
\affiliation{ 
  CAS Center For Excellence in Quantum Information and Quantum Physics, University of Science and Technology of China, Hefei 230026, China
}%
\affiliation{ 
  Hefei National Laboratory, University of Science and Technology of China, Hefei 230088, China
}

\author{Rui-Rui Li \CJKfamily{gbsn}(李睿睿)}
\affiliation{ 
  CAS Key Laboratory of Quantum Information, University of Science and Technology of China, Hefei 230026, China
}%
\affiliation{ 
  CAS Center For Excellence in Quantum Information and Quantum Physics, University of Science and Technology of China, Hefei 230026, China
}
\author{Ran He \CJKfamily{gbsn}(贺冉)}
\affiliation{ 
 Unitary Quantum Co., Ltd., Hefei, China	
}
\author{Shu-Qian Chen \CJKfamily{gbsn}(陈树谦)}
\affiliation{ 
  CAS Key Laboratory of Quantum Information, University of Science and Technology of China, Hefei 230026, China
}
\affiliation{ 
  CAS Center For Excellence in Quantum Information and Quantum Physics, University of Science and Technology of China, Hefei 230026, China
}
\author{Wen-Hao Qi \CJKfamily{gbsn}(亓文昊)}
\affiliation{ 
  CAS Key Laboratory of Quantum Information, University of Science and Technology of China, Hefei 230026, China
}
\affiliation{ 
  CAS Center For Excellence in Quantum Information and Quantum Physics, University of Science and Technology of China, Hefei 230026, China
}
\author{Jin-Ming Cui \CJKfamily{gbsn}(崔金明)}
\email[Corresponding author.\\]{jmcui@ustc.edu.cn}
\affiliation{ 
  CAS Key Laboratory of Quantum Information, University of Science and Technology of China, Hefei 230026, China
}
\affiliation{ 
  CAS Center For Excellence in Quantum Information and Quantum Physics, University of Science and Technology of China, Hefei 230026, China
}
\affiliation{ 
  Hefei National Laboratory, University of Science and Technology of China, Hefei 230088, China
}
\author{Yun-Feng Huang \CJKfamily{gbsn}(黄运锋)}
\email[Corresponding author.\\]{hyf@ustc.edu.cn}
\affiliation{ 
  CAS Key Laboratory of Quantum Information, University of Science and Technology of China, Hefei 230026, China
}
\affiliation{ 
  CAS Center For Excellence in Quantum Information and Quantum Physics, University of Science and Technology of China, Hefei 230026, China
}
\affiliation{ 
  Hefei National Laboratory, University of Science and Technology of China, Hefei 230088, China
}
\author{Chuan-Feng Li \CJKfamily{gbsn}(李传锋)}
\affiliation{ 
  CAS Key Laboratory of Quantum Information, University of Science and Technology of China, Hefei 230026, China
}
\affiliation{ 
  CAS Center For Excellence in Quantum Information and Quantum Physics, University of Science and Technology of China, Hefei 230026, China
}
\affiliation{ 
  Hefei National Laboratory, University of Science and Technology of China, Hefei 230088, China
}
\author{Guang-Can Guo \CJKfamily{gbsn}(郭光灿)}
\affiliation{ 
  CAS Key Laboratory of Quantum Information, University of Science and Technology of China, Hefei 230026, China
}
\affiliation{ 
  CAS Center For Excellence in Quantum Information and Quantum Physics, University of Science and Technology of China, Hefei 230026, China
}
\affiliation{ 
  Hefei National Laboratory, University of Science and Technology of China, Hefei 230088, China
}

\maketitle
\end{CJK*} 

\date{\today}

\begin{abstract}

Large-scale programmable trapped ion hardware, featuring high gate fidelity and long coherence times, 
is promising for realizing a practical fault-tolerant quantum computer (FTQC).
However, individual addressing (IA) methods, which are important for implementing programmable gates in near-term quantum devices, 
can lead to undesired errors between the target ions and neighboring ions.
In this work, we present a low-crosstalk optical addressing system based on 
multiple optical objectives and acousto-optic deflectors (AODs) with a symmetrical configuration.
Two counter-propagating Raman operation beams are both tightly focused, generating an overlapping spot with a waist radius of approximately $1~\upmu\mathrm{m}$, to address the target ion.
As a result, IA crosstalk, characterized by Rabi rate crosstalk on the spectator ion, is measured to be  $1.19(5)\times10^{-3}$, with the two ions separated by approximately 5.5~$\upmu\mathrm{m}$.
This low-crosstalk optical addressing system holds promise for high-fidelity entangling operations, 
and the symmetrically-configured AODs in our method can be readily extended to two dimensions to address a two-dimensional ion crystal. 
\end{abstract}

\pacs{Valid PACS appear here}
\maketitle

\section{Introduction}

Trapped ion systems, featuring high fidelity~\cite{bestSPAM,Highfidelity2014,Highfidelity2021} and long coherence times~\cite{longT2time2021}, are promising platforms for quantum computing.
Programmable two-qubit entangling gates with high fidelity are fundamental building blocks for achieving fault-tolerant quantum computing (FTQC)~\cite{FTQC1,FTQC2,QEC4,ryan1863implementing} and constructing a scalable and programmable quantum computer,
as FTQC requires the fidelity of gates to surpass the threshold of error-correcting codes~\cite{kitaev1997quantum,preskill1998reliable,knill1998resilient,aharonov1999,steane2003overhead,knill2005quantum,colorcode,surfacecode,LDPCcode}.
While high quality two-qubit gates have been demonstrated ~\cite{Highfidelity2016,Highfidelity2016_2,Highfidelity2021,loschnauer2024scalable},
extending them to quantum computation with a large ion array requires the addressing ability to perform programmability.
For instance, in the recently proposed $omg$ paradigm~\cite{allcock2021omg}, which is considered as a promising quantum computing architecture,
the manipulation of ion crystals with site distances on the order of 5~$\upmu\mathrm{m}$ (one-dimensional ion chain or two-dimensional ion crystal~\cite{kiesenhofer2023controlling, guo2024site}) requires tightly focused optical beam spots to manipulate ions at each site,
which is referred to as the optical individual addressing (IA) method.
In this case, the IA method can introduce significant crosstalk errors when applying single- and two-qubit gates, caused by residual laser spot illumination on neighboring ions.
These errors can propagate along the quantum circuit and potentially violate the fault-tolerant requirement~\cite{fang2022crosstalk,parrado2021crosstalk,Crosstalk2}, particularly in situations involving a large number of qubits. 
Therefore, developing an IA method with low crosstalk on a hardware level is important for implementing high fidelity programmable gates with the $omg$ protocol, 
thus advancing the realization of practical FTQC.

\begin{figure}[!htbp]
  \centering
    \includegraphics[width=1\linewidth]{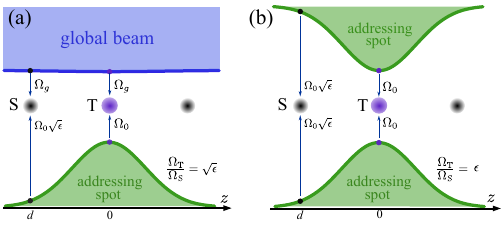}
  \caption{ Schematic diagram illustrating the comparison of Rabi rate crosstalk between two different addressing methods during Raman operation. 
  (a) The single-side addressing (SSA) method involves an addressing spot on the target ion T and a global beam on all ions. 
  (b) The double-side addressing (DSA) method involves two addressing spots from different sides overlapped on T.
  Assuming the intensity ratio of S to T is $\epsilon$ and $\epsilon \ll 1$ under a specific addressing spot, 
  then the Rabi rate crosstalk  (characterized by $\Omega_S/\Omega_T$)
  are $ \sqrt{\epsilon}$ for SSA and $\epsilon$ for DSA, respectively 
  (i.e., the crosstalk of DSA is  $\sqrt{\epsilon}$ lower than that of the SSA).
  }
  \label{f1}
\end{figure}

The IA crosstalk error induced by the intensity spillover to the neighboring ions can be characterized by the undesired Rabi rate crosstalk, which is defined as the ratio of the Rabi frequency of the spectator ion to that of the target ion, 
namely, $\varOmega _{\mathrm{S}}/\varOmega _{\mathrm{T}}$.
Tab.~\ref{tab:table1} provides a comparison of this work and other four previous predominant IA systems: multiple-channels acousto-optic modulators (MCAOMs), guided-light individual addressing system (GLIAS), acousto-optic deflectors(AODs) and micro-electromechanical systems (MEMS).
Fig.~\ref{f1} illustrates two different IA schemes with two counter-propagating laser beams to drive hyperfine atomic qubits via stimulated Raman transition (SRT) ~\cite{SRT1,SRT2,SRTion1,SRTion2}. 
\begin{table}[!htbp]
  \caption{\label{tab:table1}Comparison of various IA systems.
  Here "NA" indicates the numerical aperture of the objective used in each IA system.
  }
  \begin{ruledtabular}
  \begin{tabular}{ccccc}
  IA systems&NA&\thead{Intensity \\crosstalk} &\thead{Rabi rate \\crosstalk}&\thead{Ion \\spacing}\\
  \hline
  MCAOMs~\cite{Egan2021} & 0.4 & 1e-4& 1\%-2\% &5~$\upmu\mathrm{m}$\\
  GLIAS~\cite{binaimotlagh2023guided}& 0.4 & 1e-4&  1\%(estimate)  &4~$\upmu\mathrm{m}$\\
  AODs~\cite{IA_AOD2}& 0.6 & 4e-6 &  0.2\% -1\% &3.5~$\upmu\mathrm{m}$(min)\\
  MEMSs~\cite{IA_MEMS2021}& 0.6 & 4e-6 & 0.2\%-0.6\%  & 5~$\upmu\mathrm{m}$\\
  \textbf{This work} & \textbf{0.4$\times$2} & \textbf{$<$1e-3} &  \textbf{0.12\%} & \textbf{5.5~$\upmu\mathrm{m}$}
  \end{tabular}
  \end{ruledtabular}
\end{table}

In the single-side addressing (SSA) scheme, as depicted in Fig.~\ref{f1}(a) , the targeted ion T is illuminated by one tightly focused IA spot and one global spot, where the latter is shared by the spectator ion S, such that the IA crosstalk is proportional to the square root of the intensity crosstalk of IA spots. The lowest Rabi rate crosstalk error that the SSA scheme has ever reported to date is $1.3\times 10^{-4}$~\cite{IA_MEMS2014} (with an ion spacing of 7.4~$\upmu\mathrm{m}$), though it increases to $2\times 10^{-3}$ when implementing entangling gates (with an ion spacing of 5~$\upmu\mathrm{m}$)~\cite{IA_MEMS2021}. In the double-side addressing (DSA) scheme, as depicted in Fig. ~\ref{f1}(b), the T ion is overlapped by two tightly focused IA spots simultaneously. This allows for a substantial reduction compared to the SSA scheme, considering that the Rabi rate of the S ion scales linearly with the intensity crosstalk.

Addressing systems with fixed equal spacing IA beams, such as MCAOMs, are suitable for aligning with inhomogeneously spaced ion qubits,
while the AOD method allows for flexible alignment of the addressing beams with inhomogeneously spaced ions by fine-tuning the frequency of the driving signal.
However, the previous use of cross-configured AODs~\cite{IA_AOD2} results in additional beams that are distributed above and below the ion chain with nonzero frequency shifts, 
hindering its application in two-dimensional ion crystals. 
Moreover, the number of additional beams increases with the number of tones in the  AOD's driving signal, which compromises the effective power of the addressing laser.

To address the limitations of the cross-configured AODs scheme and take advantage of the strengths of the DSA scheme, 
we propose a low-crosstalk DSA system based on symmetrically-configured AODs in this work. 
We also note that the IonQ Forte system utilizes a similar technique, published shortly after this manuscript was submitted~\cite{chen2023benchmarking}.
Here, we employ this DSA system to manipulate the hyperfine qubit of ${^{171}}\mathrm{Yb}^+$.
We design a symmetrical optical configuration for SRT beams, 
where the orientations of two AODs, both using the ~-1st diffraction order, coincide with the axial direction to cancel the frequency shift.
By measuring the Rabi frequency of the spectator ion and the target ion, the IA crosstalk is determined to be at a level of $1.0\times10^{-3}$ between the neighboring ion sites.
This low-crosstalk DSA system, with high manipulation flexibility and potential for two-dimensional AOD addressing, holds promise as a platform for implementing quantum circuit in a large-scale programmable manner, thereby promoting the practical application of FTQC.

\renewcommand{\dblfloatpagefraction}{.6}

\section{System Setup and Characterization}

\begin{figure*}[!htbp]
  \centering
    \includegraphics[width=1.0\linewidth,scale=1.0]{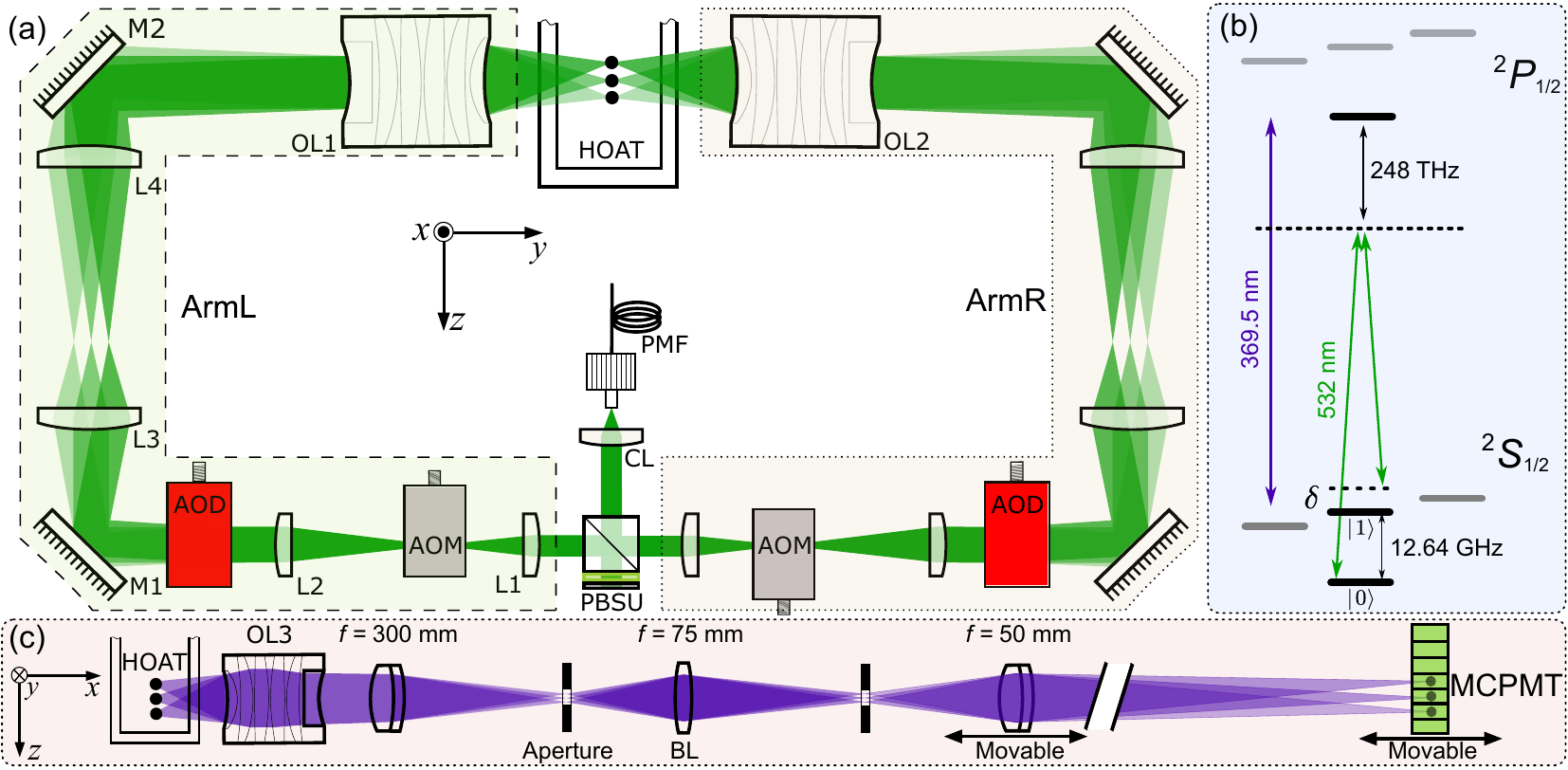}
  \caption{(a) Schematic of the optical setup of our addressing system. 
  For simplicity, two symmetrically configured arms of the interferometer are denoted as ArmL and ArmR, respectively.
  (b) Energy level schematic of ${^{171}}\mathrm{Yb}^+$ qubits. The transition wavelength for Doppler cooling is 369.5~nm, which is indicated by the purple lines. 
  The energy levels associated with the SRT are indicated by the green lines.
  (c) Schematic of the optical setup of the imaging system. 
  The distance between the last achromatic lens and the MCPMT can be adjusted to vary the magnification from 200 to 1000 times.
  HOAT: High optical access trap, PBSU: Polarizing beam splitter unit, CL: Collimating lens, 
  PMF: Polarization-maintaining fiber, AOM: Acousto-optic modulator, AOD: Acousto-optic deflector, 
  OL: objective lens, BL: Biconvex lens, MCPMT: Multi-channel photomultiplier.}
  \label{f2}
\end{figure*}

The optical setup of our experimental apparatus is depicted in Fig.~\ref{f2}(a).
For our experiment, the ${^{171}}\mathrm{Yb}^+$ qubits are confined in a Paul trap with high optical access in multiple directions~\cite{glasstrap}.
The Raman operation laser used in our experiment is a continuous wave (CW) laser modulated by a fiber electro-optical modulator (FEOM). Its wavelength is 532~nm, falling within the visible domain.
This visible Raman operation laser is delivered to a polarizing beam splitter unit (PBSU) through a single-mode and polarization-maintaining fiber and then split into two beams. 
The PBSU consists of a half-wave plate, a polarizing beam splitter, a quarter-wave plate, and a mirror.
The two portions of the Raman operation laser propagate through two symmetrically configured optical paths and are then recombined at the ion position, 
forming a Mach-Zehnder interferometer.

We designate the two arms of the interferometer as ArmL and ArmR for simplicity.  
The optical configuration of ArmL is illustrated as an example, considering the symmetrical nature of the interferometer.
The Raman operation beam is initially magnified by a pair of achromatic lenses (L1 and L2).
The acousto-optic modulator (AOM, Gooch \& Housego, 3200-121) placed between the two lenses serves the purpose of shifting the relative frequency of two counter-propagating Raman operation lasers~\cite{Li:22}.
Unlike the cross-configuration of AODs~\cite{IA_AOD2} used in the SSA method, 
AODs (AA Opto Electronic, DTSX-400) here are placed on ArmL and ArmR respectively, with the string of diffracted beams of the AOD aligned along the direction of the ion chain.
The diffracted beams from the AOD are then reflected by two mirrors (M1 and M2) and magnified by the second pair of lenses (L3 and L4).
Finally, the beams are focused onto the plane of ions at addressing spots by a 0.4~NA objective lens (OL1).
The diffracted beams from the two arms with the same frequency shift will illuminate the same ion, 
canceling the position-dependent frequency shift,
provided that the optical configuration is symmetrical.

The axis of the imaging system is perpendicular to the plane of the Raman operation beam, as shown in Fig.~\ref{f2}(c). 
Additionally, a 0.4~NA objective lens (OL3) is used to collect the fluorescence emitted by the ions. 
The multi-channel photomultiplier (MCPMT) consists of 32 segmented channels,
with each channel spaced 200~$\upmu\mathrm{m}$ apart and with a width of 800~$\upmu\mathrm{m}$.
This configuration allows for the parallel detection of the states of multiple ions, as the fluorescence from each ion is directed to the corresponding MCPMT channels.
To minimize the MCPMT crosstalk originating from the intensity crosstalk of the ion image between MCPMT channels, achromatic lenses and apertures are used to mitigate optical aberrations.

\begin{figure}
  \centering
    \includegraphics[width=1.0\linewidth,scale=1.0]{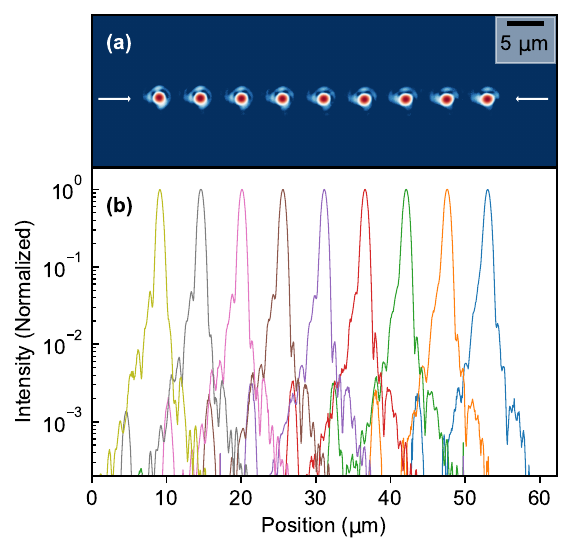}
  \caption{	
	(a) Image of diffracted IA spots of the AOD at the ion plane. 
	The distance between IA spots corresponds the distance between the ions. 
	(b) The profiles of the IA spots in the cross-section. 
	The shape of spots deviates from a Gaussian distribution primarily due to the aberrations caused by the homemade objective.
  }
  \label{f3}
\end{figure}

To characterize the intensity crosstalk between multiple IA spots diffracted from the AOD, 
the laser beams focused on the ion plane are re-imaged onto a high dynamic range camera. 
The AOD generates a single diffracted IA spot by applying a monochromatic drive signal. 
Then, we sequentially increase the driving signal frequency to obtain nine diffracted spots.
The array of driving frequencies here correspond to those applied to the AOD when addressing multiple ions.
We then stitch together the images of each diffracted IA spot, as shown in Fig.~\ref{f3}(a).
The arrows in Fig.~\ref{f3}(a) indicate the orientation of the ion chain and define a cross-section across the center of the IA spots.
Fig.~\ref{f3}(b) shows the profiles of the IA spots in this cross-section.
From Fig.~\ref{f3}(b), it can be observed that the IA spot crosstalk of the neighboring sites of ions is at the level of $1.0\times10^{-3}$,
indicating the Rabi rate crosstalk error will be at the same level.

\section{Experimental Details and Results}

The schematic diagram of the energy levels of the ${^{171}}\mathrm{Yb}^+$ ion is shown in Fig.~\ref{f2}(b).
The qubits are defined by the two hyperfine ground states of ${^{171}}\mathrm{Yb}^+$, namely,
$|0\rangle \equiv {^{2}}S_{1/2}|F=0,\;m_F=0\rangle$ and $|1\rangle \equiv {^{2}}S_{1/2}|F=1,\;m_F=0\rangle$. 
Doppler cooling, electromagnetic induced transparency (EIT) cooling~\cite{EITcooling1,EITcooling2}, 
optical pumping, and state-dependent fluorescence detection are achieved using a 369.5~nm CW laser. 
Repumping from the ${^{2}}D_{3/2}$ state is accomplished using a 935~nm CW laser~\cite{firstYb}.
The qubits are driven by the SRT using two coherent laser fields with a frequency gap 
equal to the frequency splitting of the hyperfine ground states, i.e., $\omega_\mathrm{HF}=$12.64~GHz.

\begin{figure}
  \centering
    \includegraphics[width=1.0\linewidth,scale=1.0]{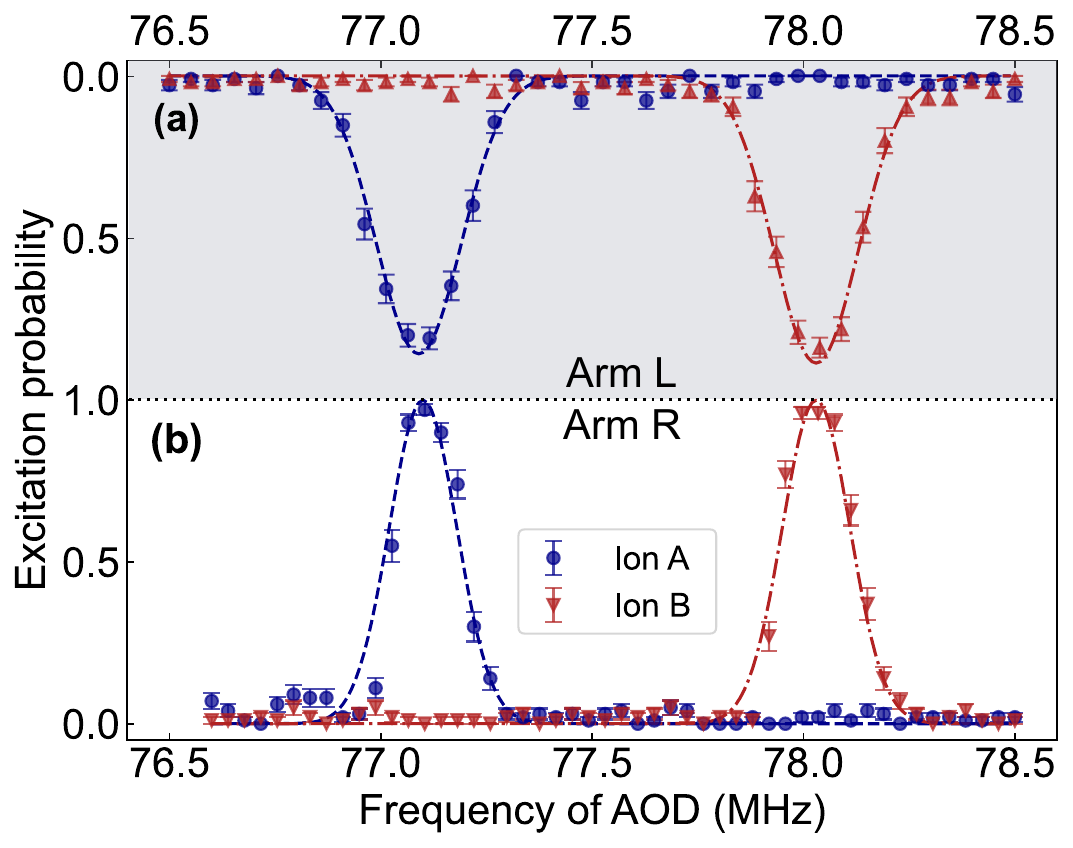}
  \caption{The excitation probability of ion A and ion B when scan the IA spot over the ion chain. 
	For ArmL (a) and ArmR (b), we sweep the driving frequencies of AODs respectively so that the position of the IA spot sweeps across two ions. 
	The dashed lines are the Gaussian fits to the data from the corresponding channels. 
	All error bars indicate $68\%$ confidence intervals (based on the Gaussian distribution).
  }
  \label{f4}
\end{figure}

For the first experiment, we carefully align the diffracted spots of both AODs with the ions and verify the symmetry of the optical configuration. 
In the AOD-based IA system, each specific diffracted spot of the AODs corresponds to its respective addressed ion.
We initially confine two ions in the trap, denoted as ion A and ion B, respectively.
After tuning the optics, we sweep the drive frequency of the AODs, and accordingly, the diffracted IA spot moves along the orientation of the ion chain. 
In this experiment, the power and duration of the Raman operation laser are both fixed, 
so the probability of the ion being driven to the $|1\rangle$ state depends on the overlapping area with the scanning spot.
A measurement result of the Gaussian distribution curve is shown in Fig.~\ref{f4}.
Each data point of Fig.~\ref{f4} is an average of 100 shots. 
The experimental sequence consists of 1~ms Doppler cooling, 1~ms EIT cooling, 20~$\upmu\mathrm{s}$ pumping, Raman operation laser with a fixed interaction time, and 1~ms detecting.

We utilize the function 
\begin{equation}
    A\left( x \right) =\sum_{i}{A_i e^{-2\left( \frac{x-x_i}{w_i} \right)^2}}
    \label{e1}
\end{equation}
to fit the data, where $i$ denotes ion A or ion B, $A_i$ represents the maximum excitation probability, 
$w_i$ represents the waist of each IA spot,
$x_i$ denotes the center position of the IA spot, 
and $x$ is the positional parameter of the scanning spot.
The positional parameters have been converted into the driving frequency of the AOD.
The centers of ion A and ion B are thus fitted as 77.079(4)~MHz and 78.027(4)~MHz, respectively.
Given that the distance between ion A and ion B is approximately 5.5~$\upmu\mathrm{m}$, 
the waists of IA spots from ArmL and ArmR are calculated as 1.15(3)~$\upmu\mathrm{m}$ and 0.93(3)~$\upmu\mathrm{m}$ respectively, 
which are close to the theoretical diffraction limit (0.81~$\upmu\mathrm{m}$).
Moreover, the scan results confirm the symmetry of the optical configuration, namely, the diffracted IA spots with identical frequency shifts illuminate the same ion.

\begin{figure*}[!htbp]
  \centering
    \includegraphics[width=1.0\linewidth,scale=1.0]{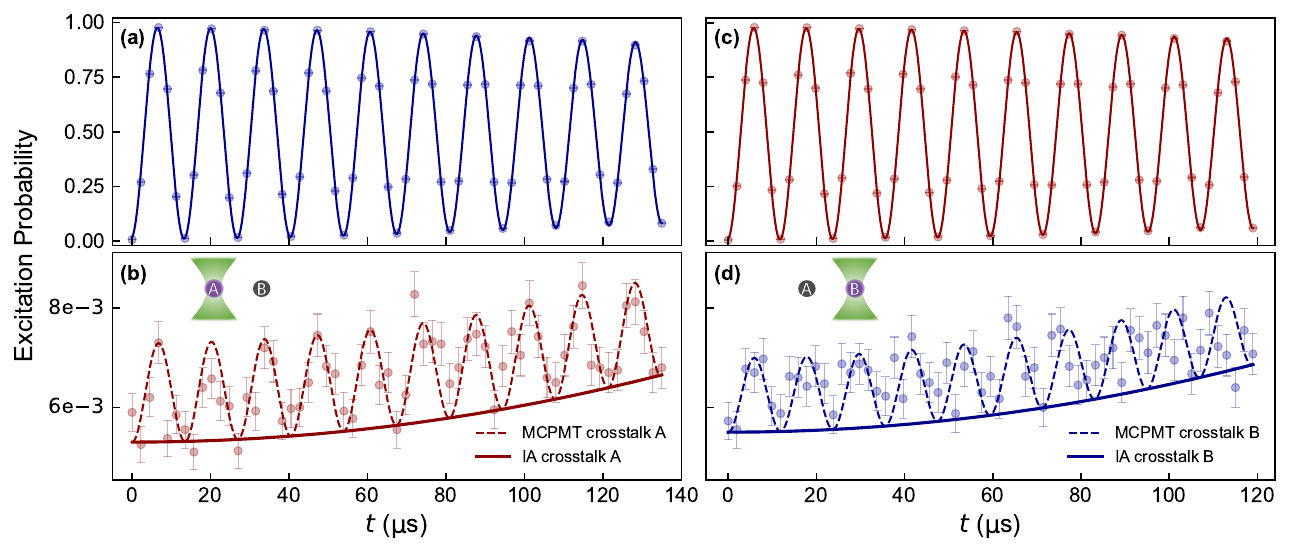}
	\caption{
	We initially confine two ions, ion A and ion B.
	Ion A is individually addressed as the target, and the Rabi oscillation is illustrated in (a).
	The Rabi frequency of ion A is determined to be $2 \pi \times 74.14$~kHz.
	We detect ion B at the same time, and the result is shown in (b).
	There are three main sources contributing to the excitation probability of ion B: the state detecting crosstalk error induced by the MCPMT, the single qubit state preparation and measurement (SPAM) error,  and the Rabi oscillation driven by IA crosstalk (details in Appendix A).
	The dashed line shows the sum of IA crosstalk, SPAM error, and MCPMT crosstalk, of which the oscillation synchronizes with the target ion A.
	The solid line shows the sum of IA crosstalk and SPAM error.
	From Eq.~\ref{e2}, the Rabi frequency of ion B is calculated to be  $2 \pi \times 88(4)$~Hz.
	(c) and (d) represent the situation where ion B is the target ion and ion A is the spectator ion.
	The Rabi frequencies are $2 \pi \times 84.13$~kHz and $2\pi\times100(6)$~Hz, respectively.
	All error bars indicate $68\%$ confidence intervals (based on the Gaussian distribution).
   }
  \label{f5}
\end{figure*}

For the subsequent experiment, we investigate the Rabi rate crosstalk when addressing ion A or ion B individually.
We set the drive frequency of the AODs as the corresponding frequency of ion A or ion B, specifically 77.08~MHz and 78.03~MHz.
To eliminate unwanted Rabi drive due to pulse rise/fall times, where the frequency bandwidth of the pulse is converted into the spatial extent of the beam by the AOD, the AOD is initially activated for a duration of 2~$\upmu\mathrm{s}$ to prepare before opening the AOM to allow the laser to pass through the modulators and address the ion.

The Rabi oscillations of ion A and ion B versus interaction time are recorded in Fig.~\ref{f5}.
Each data point is an average of 40000 shots.
The experimental sequence consists of 1~ms Doppler cooling, 1~ms EIT cooling, 20~$\upmu\mathrm{s}$ pumping,~$t$ Raman operation, and 1~ms detecting.
For the target ion, the Rabi frequencies of ion A and ion B are fitting to be $2 \pi \times 74.14$~kHz and $2 \pi \times 84.13$~kHz respectively, as shown in Fig.~\ref{f5}(a) and (c).
For the spectator ion, there are three main sources contributing to the excitation probability recorded by the MCPMT: the state detecting crosstalk error induced by the MCPMT, the single qubit state preparation and measurement (SPAM) error, and the Rabi oscillation driven by IA crosstalk, which is the main focus here. 
The excitation probability of the spectator ion is well described by Eq.~\ref{e2} (see~{Appendix A}), derived from which the Rabi frequencies of spectator ion A and ion B are determined to be $2 \pi \times 88(4)$~Hz and $2\pi\times100(6)$~Hz respectively, as shown in Fig.~\ref{f5}(b) and (d). 
Therefore, the Rabi rate crosstalk on ion A and ion B are calculated to be $1.19(5)\times10^{-3}$ and $1.19(7)\times10^{-3}$, respectively.

\section{Discussion and Outlook}
In this work, we demonstrate a low-crosstalk optical double-side addressing system based on symmetrically-configured AODs.
The use of AODs allows for flexible alignment of the laser beams to the ion spots individually, without the need to move ions. 
By employing a double-side symmetrical configuration, we are able to achieve a low crosstalk between ions at a level of $0.1\%$,
which is consistent with the intensity of spots spillover crosstalk.
Moreover, we can mitigate the crosstalk, by improving the homemade objective, or using a higher NA objective (e.g. 0.66 NA~\cite{glasstrap}).
Furthermore, by using spatial light modulators to calibrate the aberration of the optical system of trapped ions ~\cite{qian2021super, zupancic2016ultra,shih2021reprogrammable}, an IA crosstalk of $<1.0\times10^{-4}$ could be achieved in our addressing system. 
It is worth noting that large scale quantum simulation has been successfully achieved on the 2D crystal~\cite{guo2024site}, and our method offers a potential ability for quantum computation by replacing the 1D-AOD with a 2D-AOD on both sides.
\\
\\

\section*{Acknowledgements}
This work was supported by the National Key Research and Development Program of China (No. 2017YFA0304100), 
the National Natural Science Foundation of China (Grant No. 11774335, No. 11734015, and No. 12204455), 
the Key Research Program of Frontier Sciences, CAS (Grant No. QYZDY-SSWSLH003), 
and the Innovation Program for Quantum Science and Technology (Grant No. 2021ZD0301604 and No. 2021ZD0301200). 
The HOAT used in the work was fabricated at the USTC Center for Micro and Nanoscale Research and Fabrication.

Yi-Long Chen and Rui-Rui Li contributed equally to this work.

\providecommand{\noopsort}[1]{}\providecommand{\singleletter}[1]{#1}%

\section*{Appendix A}
\setcounter{figure}{0}
\setcounter{table}{0}
\renewcommand{\thetable}{A\arabic{table}}
\renewcommand{\thefigure}{A\arabic{figure}}
 
To accurately determine the IA crosstalk, it is necessary to carefully consider the sources that contribute to the bright counts of the spectator ion.
After a Rabi oscillating operation of time $t$, ideally, the detecting laser will cause both the spectator ion and the target ion to collapse into the state  $|i\rangle\otimes|j\rangle$ with a probability $P(|i\rangle|j\rangle,t)=P_{S}(t)P_{T}(t)$.
Here, i = (0,\;1) denotes the (dark, bright) state respectively and so is j, and $P_{S/T}(t) = \frac{1}{2}\big( 1-\cos ( \varOmega _{S/T}t )\big)$, where $\varOmega _{S/T}$ refers to the Rabi frequency of the spectator/target ion. However, the probability of the final readout of the spectator ion being in state $|1\rangle$ given the condition that $|S\rangle\otimes|T\rangle=|i\rangle\otimes|j\rangle$, denoted as $P\big({S =|1\rangle\;\Big|\;|i\rangle|j\rangle}\big)$, will be of more complex.

The spectator ion will be determined to be in state $|1\rangle$, whenever the corresponding channels of MCPMT collect photons scattered either from the spectator ion itself or the spectator ion induced by MCPMT crosstalk.
If the target ion is in state $|0\rangle$, the number of spillover photons is too small to be considered, so the spectator ion will only be subjected to the SPAM error.
The probability of the final readout of the spectator ion being in state $|1\rangle$ is therefore $\epsilon_{01}$ if $|S\rangle = |0\rangle$ and $1-\epsilon_{10}$ if $|S\rangle = |1\rangle$, where $\epsilon_{01}$ and $\epsilon_{10}$ represent the single qubit SPAM error.
If the target ion is in state $|1\rangle$, we assume an MCPMT crosstalk $\epsilon_{MP}$, which is the probability that spillover photons alone would cause the spectator ion to be read as state $|1\rangle$. In this case, the probability of the readout of the spectator ion being in state $|1\rangle$ is $1-(1-\epsilon_{01})(1-\epsilon_{MP})$ if $|S\rangle = |0\rangle$ and $1-\epsilon_{10}(1-\epsilon_{MP})$ if $|S\rangle = |1\rangle$. 
These results are all listed in Tab.~\ref{tab:table2}.

\begin{table}[!htbp]
	\renewcommand{\arraystretch}{1.2}
  \caption{\label{tab:table2}Probability of various cases. 
  }
  \begin{ruledtabular}
  \begin{tabular}{ccccc}
  $|S\rangle\otimes|T\rangle$ & $P\big({|i\rangle|j\rangle},t\big)$ & $P\big({S =|1\rangle\;\Big|\;|i\rangle|j\rangle}\big)$ \\
  \hline
  ${|0\rangle\otimes|0\rangle}$ & $\big(1-P_S(t)\big)\big(1-P_T(t)\big)$ & $\epsilon_{01}$\\
  ${|0\rangle\otimes|1\rangle}$ & $\big(1-P_S(t)\big)P_T(t)$     & $1-(1-\epsilon_{01})(1-\epsilon_{MP})$\\
  ${|1\rangle\otimes|0\rangle}$ & $P_S(t)\big(1-P_T(t)\big)$     & $1-\epsilon_{10}$\\
  ${|1\rangle\otimes|1\rangle}$ & $P_S(t)P_T(t)$         & $1-\epsilon_{10}(1-\epsilon_{MP})$\\
  \end{tabular}
  \end{ruledtabular}
\end{table}

As a result, the probability of the form is obtained by summing all the cases:
\begin{equation}
\begin{aligned}
	P\big({S=|1\rangle},t\big) = &\sum_{i,j}P\big({S =|1\rangle\;\Big|\;|i\rangle|j\rangle}\big){P\big({|i\rangle|j\rangle},t\big)}
\\	\approx &\underbrace{\left\{\frac{\epsilon_{MP}}{2}\big(1+\cos \left( \varOmega _{S}t \right)\big)\times\frac{1}{2}\big( 1-\cos \left( \varOmega _{T}t \right)\big)\right\}}_{\displaystyle \rm{term \;1}}
\\	+ &\underbrace{\left\{\big(1-\big(\epsilon_{01}+\epsilon_{10}\big)\big)\times\frac{1}{2}\big( 1-\cos \left( \varOmega _{S}t \right)\big)\right\}}_{\displaystyle \rm{term \;2}}
\\	+ &\epsilon_{01}
	\label{e2}
\\ 
\end{aligned}
\end{equation}
The overall measured crosstalk consists of three terms.
Term 1 represents the state detecting crosstalk error induced by the MCPMT, behaving like a fast wave synchronized with the target ion. Term 2 shows the Rabi oscillation driven by IA crosstalk. The last term, SPAM error term $\epsilon_{01}$, elevates the overall excitation probability curve to a baseline. 
$\epsilon_{10}$, $\epsilon_{01}$, and $\epsilon_{MP}$ in our experiment are approximately 1.10$\%$, 0.53$\%$, and 0.20$\%$ for ion A and 1.10$\%$, 0.55$\%$, and 0.15$\%$ for ion B.
These values were obtained using the histogram method~\cite{debnath2016programmable}. 
The slight asymmetry of crosstalk is due to the asymmetry of the MCPMT and the imaging system.
\\

\section*{Appendix B}
\setcounter{figure}{0}
\renewcommand{\thetable}{B\arabic{table}}
\renewcommand{\thefigure}{B\arabic{figure}}

\begin{figure}[!htbp]
  \centering
      \includegraphics[width=1.0\linewidth,scale=1.0]{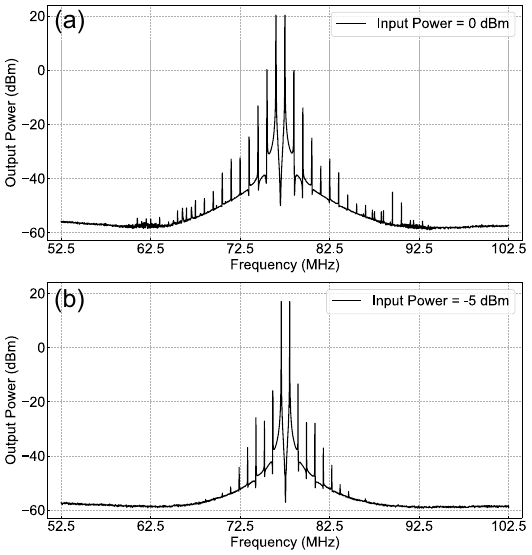}
  \caption{The output spectrum of the RF amplifier when the input power is 0~dBm in (a) and -5~dBm in (b).}
  \label{figB1}
\end{figure}

\begin{figure}[!htbp]
  \centering
      \includegraphics[width=0.9\linewidth,scale=1.0]{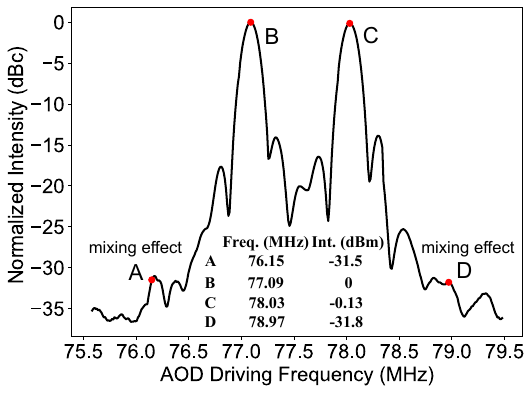}
	\caption{The intensity distribution of diffracted spots of AOD when the input power is -5~dBm.
	Spot B and C are two desired tones and spot A and D correspond to two third-order IMD tones.
	The embedded table shows the frequencies and intensities of spots.
   }
  \label{figB2}
\end{figure}

The intermodulation distortion (IMD) induced by multiple tones of the input RF signal when multiple ions are addressed simultaneously will cause extra crosstalk.
A simple method to mitigate IMD is to reduce the power of the input signal applied to the RF amplifier.
In our experiment, we apply two tones, 77.09~MHz and 78.03~MHz, to the RF amplifier and then 
measure the output spectrum with the input power of 0~dBm and -5~dBm, respectively, as shown in Fig.~\ref{figB1}.
After lowering the input power, the power of third-order IMD components is about 30~dB lower than the desired tones.
We then apply this RF signal to the AODs and measure the intensity distribution of diffracted spots, as shown in Fig.~\ref{figB2}.
The intensities of the spots A and D, caused by third-order IMD, are -31.5~dBc and -31.8~dBc, respectively.
It can be deduced that the crosstalk caused by IMD are $7.07 \times 10^{-4}$ and $6.61 \times 10^{-4}$, which are comparable to the IA crosstalk at a level of $0.1\%$. If IMD could be further suppressed (e.g., using an RF amplifier with higher 3rd order intercept point), the IA crosstalk would be lower.

\end{document}